\documentstyle[aps,prl]{revtex}
\draft
\input{psfig}
\begin{document}
\twocolumn[\hsize\textwidth\columnwidth\hsize\csname@twocolumnfalse\endcsname
\title{Stretching of vortex lines 
and generation of vorticity in the three-dimensional  
complex Ginzburg-Landau Equation} 
\author{I. S. Aranson$^{1,2}$ and A. R. Bishop$^3$ } 
\address{$^1$ Department of Physics, 
Bar Ilan University, Ramat Gan 52900, Israel\\ 
$^2$ Argonne National Laboratory, 
9700 South  Cass Avenue, Argonne, IL 60439\\
$^3$ Theoretical Division and Center for Nonlinear Studies, \\
Los Alamos National Laboratory, Los Alamos, NM 87545} 
\date{\today} 
\maketitle 
\begin{abstract} 
The dynamics of curved vortex filaments 
is studied analytically and numerically in the framework of  
a three-dimensional 
complex Ginzburg-Landau equation (CGLE). 
It is proved 
that a straight vortex line is unstable with respect to 
spontaneous stretching and bending
in a certain range
of parameters of the CGLE, resulting in formation of persistent
entangled vortex configurations. 
The analysis 
shows that the standard  approach relating the 
velocity of the filament with the local curvature, is 
insufficient to describe the instability and self-generation of vorticity. 
\end{abstract} 
\vskip1pc] 
\narrowtext 
The complex Ginzburg-Landau equation (CGLE), 
derived some 20 years ago by Newell 
\cite{newell} and Kuramoto \cite{kuramoto}
has become a paradigm model 
for a qualitative description of weakly-nonlinear oscillatory  
media (see for review \cite{cross}). Under appropriate scaling of 
the physical variables, the equation assumes the universal form  
\begin{equation} 
\partial_t A=A-(1+ic)|A|^2A+(1+ib) \Delta A,  
\label{cgle} 
\end{equation} 
where $A$ is the complex amplitude, $b$ and $c$ are real parameters, 
and $\Delta=\partial_x^2+\partial_y^2+\partial_z^2$ is a three-dimensional 
Laplace operator.  
Although the equation is formally valid only at the threshold of  
a supercritical Hopf bifurcation, it has been found that the CGLE
often reproduces qualitatively  
correct phenomenology over a much wider range of the parameters. As a  
results, the predictions drawn from the analysis of the CGLE (mostly  
in one and 
two spatial dimensions, see e.g. \cite{akw,akw2,chate}) were  
recently successfully confirmed by experiments in optical and 
chemical systems \cite{arrechi,flessel}. 
Moreover, some results obtained from the CGLE (for example, symmetry breaking 
of spiral pairs) was instructive for interpretation of  
experiments in far more complicated systems of 
chemical waves \cite{perez} and colonies of amoebae 
\cite{alt,gold}. 
 
Recently, the dynamics of {\it three-dimensional}  (3D) vortex filaments has
attracted  
substantial attention \cite{frisch,gog,karma,winfree,winfree1}. 
In the context of the  
3D CGLE,  Gabbay Ott and Guzdar \cite{gog}  
applied a generalization of Keener's method for  
a scroll vortex in reaction-diffusion systems \cite{keener}. They derived  
that a vortex ring of radius $R$ collapses in  finite time 
over the  entire  
range of parameters  $(b,c)$ of Eq. (\ref{cgle}) 
according to the following evolution law 
 \begin{equation} 
\frac{dR}{dt}=-\frac{1+b^2}{R}. 
\label{ott} 
\end{equation}  
This result  
generalizes  Keener's ansatz by including the curvature-induced  
shift of the filament's wavenumber 
and corrects the erroneous estimate of Ref. \cite{frisch}. 
Thereby, as follows from Eq. (\ref{ott}), vorticity initially existing 
in the system will always decay (if, of course, there is no bulk  
instability of the waves emitted by the vortex filament),  
and under no condition can
the vortex filament expand. The persistence of vortex filaments, 
known under some conditions for reaction-diffusion systems,  
was attributed in the context of the CGLE to the effect of higher-order  
corrections to the CGLE.  
 
In this Letter we show that under very general conditions and in  
an extensive part of the parameter space vortex filaments {\it expand} 
and result in persistent vortex configurations even if there is no  
bulk instability of emitted waves. 
The condition for the expansion of the vortex 
filament is simply $b >b_c(c) \gg 1$. In this limit Eq. (\ref{ott})  
is not valid, because  formally higher-order corrections,  
omitted in Eq. (\ref{ott}), cause severe instability of the filament  
and persistent  stretching. This instability is a three-dimensional 
manifestation of the two-dimensional core instability of spiral waves
(called in Ref. \cite{akw2} {\it acceleration instability}). It
originates from  breakdown of the Galilean invariance of the CGLE for 
any $\epsilon=1/b\ne0$, causing spontaneous acceleration of  
the spiral waves for $\epsilon \le \epsilon_c$ \cite{akw2}. Although 
in 2D the instability is relatively weak (the growth rate is proportional 
to $\epsilon$), the situation is different in three dimensions.  
As we will show, the bending of the  
filament greatly enhances the instability, resulting in formation, after some 
transient, of a highly entangled and dense vortex configuration. 
The condition $b \gg 1$ is readily  fullfiled for many physical and  
chemical systems. For example, in the context of nonlinear optics,
where the CGLE can be  derived from the 
Maxwell-Bloch equation in the good cavity
limit \cite{lmn},  
this parameter is very small: $\epsilon=1/b \sim 10^{-4}-10^{-3}$.  
For an oscillating chemical reaction the diffusion rates  of various 
components 
can be varied over a wide range by adding extra chemicals. 
 
As a test for instability, we consider the dynamics of a weakly curved  
vortex filament. We apply the generalization of the method of Ref. \cite 
{akw2} for the case of 3D vortices, and  make perturbations near  
the 2D spiral wave solution to the 
CGLE. For convenience, we redefine 
${\bf r} \to {\bf r}/\sqrt{b}$. Then Eq. (\ref{cgle})  assumes the 
form
\begin{equation}
\partial_t A=A-(1+ic)|A|^2A+(\epsilon+i) \Delta A.
\label{cgle1}
\end{equation}
In the following discussion,  we assume $ 
0<\epsilon \ll 1$ to be a small parameter. Our objective is to relate 
the acceleration  of the vortex filament $\partial_t {\bf v}$ with  
the  velocity $  {\bf v}$ and local curvature $\kappa$ of the filament. 
We obtain that for $\epsilon \ll 1$ the equation of
filament motion can be 
written as 
\begin{equation} 
 \partial_t {\bf v} + 
 \hat K [\epsilon  {\bf v} + (1+\epsilon^2) \kappa {\bf N}]  =0, 
\label{accel} 
\end{equation} 
where ${\bf N}$ is the unit vector pointing toward the center of curvature, 
and ${\bf B}$ is the unit vector perpendicular to the filament and ${\bf N}$. 
Velocity and acceleration have correspondingly two components,  
${\bf v} = (v_N,v_B), \partial_t {\bf v}=(\partial_t v_N, \partial_t v_B)$.  
$\hat K$ is the friction 
tensor satisfying $K_{11}=K_{22}, K_{12}=-K_{21}$. Dropping the first 
term in Eq.(\ref{accel}), we reproduce the result of Ref. \cite{gog} 
for the collapse rate 
$v_N = -(1+\epsilon^2) \kappa/\epsilon$ (because $\kappa=1/R$). 
However, if $K_{11} <0$, which we will show
is the case for  $\epsilon < \epsilon_c(c) $, 
the acceleration has a significant
effect and leads to an ultimate expansion  
of the vortex filaments and the persistence of entangled vortex  
configurations. In principle, the numerical value of $ \epsilon_c(c)$
does not need  to be very small. For example, for $c=0.5$ we obtained
$\epsilon_c\approx0.165$, which corresponds to 
$b\approx 6$.

In order to develop the perturbation theory for a weakly-curved 
vortex filament in 3D, we begin with  
the 
stationary  one-armed  isolated spiral 
solution to Eqs.\ (\ref{cgle},\ref{cgle1}), which  is of the form  
\begin{eqnarray} 
A_0(r,\theta) = F(r) \exp i [ \omega t \pm \theta + \psi(r)] \;,  \label{spir} 
\end{eqnarray} 
where $(r, \theta)$ are polar coordinates. The real functions $F$ and $\psi$ 
have the following asymptotic behavior $F(r) \to \sqrt{1 - \epsilon k^2} 
\;, \psi^\prime(r) \to k $ for $r \to \infty $ and $F(r) \sim r \;, 
\psi^\prime(r) \sim r$ for $r \to 0 $. The wavenumber $k$ of the  
waves emitted by the spiral is determined uniquely  
for given $\varepsilon,c$ \cite{Hagan}.  
For $\epsilon=0$ one has a type of Galilean invariance  
and then, in addition to the stationary spiral, there exists 
a family of spirals moving with arbitrary constant velocity ${\bf v}$ 
\cite{akw2},  
\begin{equation} 
A(r ,t)= F(r^\prime) \exp i[ \omega^\prime t + \theta + \psi(r^\prime) -%
\frac{{\bf r}^\prime \cdot {\bf v}}2],  \label{mov} 
\end{equation} 
where ${\bf r}^\prime = {\bf r} + {\bf v} t$,  
$\omega^\prime= \omega +{\bf v}^2/4$, 
and the functions $F, \psi $ are those of Eq.\ (\ref{spir}). (This 
invariance holds for any stationary solution). For $\epsilon \ne 0$ the 
diffusion term $\sim \varepsilon \Delta A$ destroys the family and
leads to acceleration/deceleration
of the spiral proportional to $\epsilon v$, depending 
on the value of $\epsilon$. 
For $\epsilon<\epsilon_c(c)$ the  spiral is unstable 
with respect to spontaneous acceleration since $K_{11} <0$ \cite{akw2}. 
 
Let us now consider the dynamics of an almost straight vortex line in the 
filament-based coordinate system (see for details \cite{gog}).  
The position in space $X$ is represented by 
local coordinates $s,\tilde x, \tilde y$, where $s$ is the  
arclength along the filament, and ${\bf X} ={\bf R} 
(s)+\tilde x {\bf N}(s) + 
\tilde y {\bf B}(s)$,  
where ${\bf R}$ is the coordinate of the filament. 
In this basis the weakly curved filament moving  
with velocity ${\bf v}$ can be written in the form 
\begin{eqnarray} 
A(r ,t)& =& F(r^\prime) 
  \exp  \left[ i \left( \omega^\prime t + \theta + 
\psi(r^\prime)  
  -
 \frac{{\bf r}^\prime \cdot {\bf v}}2
\right.  \right. 
\nonumber \\
&+&
\left. \left. 
\frac{\delta k_x  
\tilde x}{2} \right) \right] 
+  W(r^\prime,\theta,s).  
 \label{mov1} 
\end{eqnarray}  
Here ${\bf r^\prime=r+vt}$,  
$W$ is the perturbation to the spiral solution which we require to be small, 
and $\delta k_x$ is the correction to the wavenumber due to curvature of  
the filament. In principle, this correction can be absorbed in $W$ but 
it is convenient to retain this form since we can cancel part of  
the perturbation exactly by adjusting  $\delta k_x$ \cite{gog}. 
The perturbation procedure to derive Eq. (\ref{accel}) is practically 
identical to that of Refs. \cite{akw,akw2}. 
Substituting the ansatz Eq. (\ref{mov1}) into the CGLE, and
assuming $|\partial_t v| , \kappa  \ll 1$, one obtains from 
Eq.\ (\ref{cgle1}) in first order in $\epsilon$ an inhomogeneous linear 
equation for the correction $W$, which is of the form 
$\hat L W  = H$, 
where $\hat L W= (1+i \omega-2(1+ic)|A_0|^2 ) W -(1+ic)A_0^2 W^* +
i \Delta_\perp W$, $\Delta_\perp=\partial_{\tilde x}^2 +
 \partial_{\tilde y}^2$.  In first order 
the $W$-independent part $H$  has two 
terms, $H=H_{tr}+H_{rot}$, where $H_{tr}$ contributes to 
the displacement of the vortex filament, and $H_{rot}$
contributes to the shift of the vortex frequency. We are interesting only
in $H_{tr}$ (the  contribution $H_{rot}$ is not 
responsible for the instability), which is of the form:
\begin{equation}
H_{tr}=- {\bf r^\prime \partial_t v} \frac{i A_0}{2}
+i\epsilon {\bf v} \nabla A_0 +(\epsilon+i)(\kappa-i  \delta k_x) \nabla A_0. 
\label{H1}
\end{equation}
The derived system of equations is very close to that considered in 
the Ref. \cite{akw2} in the context of the acceleration instability of
a spiral wave in 2D. The acceleration can be obtained 
as a result of the solvability condition requiring that $W$ be regular at the 
core of the spiral and does not diverge {\it exponentially} at large 
$r$. (Slower, power-law, divergence of $W$ is permitted). 
To impose the solvability condition  a specific  numerical procedure is 
required (see for details \cite{akw,akw2}). 
Eq. (\ref{H1}) contains 
an extra term  $(\epsilon+i)( \kappa-i \delta k_x)   \partial_{\tilde x}A_0$
 with respect to that 
considered in \cite{akw2}, 
which originates from
the Laplace operator in a local basis 
(in the limit of  
small curvature $\kappa$ and torsion $\tau$): 
$\Delta = 
 - \kappa \partial_{\tilde x} + 
\partial_{\tilde x}^2 + 
 \partial_{\tilde y}^2 +\partial_s^2 + ...$. 
However, additional numerical integration is not necessary to account for
the effect of this term, because, 
as was found in Ref. \cite{gog}, this term 
can be canceled  {\it exactly} 
to first order in $\kappa$  by a 
proper choice of $\delta k_x=-\epsilon \kappa$.  
After this modification, $H_{tr}$ assumes the form 
$H_{tr} = -{\bf r^\prime \partial_t v} \frac{i A_0}{2}  
+i\epsilon {\bf \bar v} \nabla A_0$, where $\bar v_N=v_N+(1+\epsilon^2) 
\kappa/\epsilon$ 
and $\bar v_B = v_B$. It is easy to see that it now coincides 
with the corresponding function considered in Ref. \cite{akw2}. 
Following the lines of analysis of Ref.  
\cite{akw2}, we observe that  the  relation between 
$\partial_t v$ and $\bar v$  is identical to that considered in \cite{akw2},
and is of the form  
\begin{equation}
{\bf \partial_t v} +\epsilon  \hat K {\bf \bar v}=0. 
\label{accel2}
\end{equation}  
Thus we  obtain  
Eq. (\ref{accel}). The coefficients of the friction tensor  
$K_{ij}$ coincide with those calculated in Ref. \cite{akw2}  
using numerical matching, and for $\epsilon<\epsilon_c$ it was found 
that $K_{11}<0$, which would guarantee the instability
with respect to spontaneous  
acceleration of the vortex filament. Moreover, since in
3D the direction of motion of the filament 
in general varies along the filament, this instability
results in stretching of the vortex.  
 
In the case of the acceleration instability, 
$K_{11} <0$ and the curvature has a strong destabilizing effect. Indeed, 
for an almost straight vortex parallel to the axis $z$, 
we can parameterize the position along the  
filament by the $z$ coordiante: $(X_0(z),Y_0(z))$.  
Since in this limit  
the arclength $s$ is close to $z$, 
the curvature correction to the velocity $\kappa {\bf N}$ 
is simply $\kappa {\bf N}  =-(\partial_z^2 X_{0},\partial_z^2 Y_{0})$. 
After simple algebra one derives 
the following relation for the growth rate $\lambda(k)$ of linear perturbation 
$X_0(z),Y_0(z) \sim \exp[\lambda(k) t+ i k z] $:   
 $\lambda^2+(K_{11}+i K_{12} )(
\epsilon \lambda+ k^2)=0$. For $k \gg \epsilon $ we obtain 
$\lambda \approx  
 \sqrt {-(K_{11}+i K_{12} )} k \gg \epsilon $.
Thus, for finite $k$ the growth rate $\lambda(k)$
may significantly exceed the increment of the acceleration instability in 
2D (corresponding to $k=0$): $ \lambda_0 =- \epsilon (K_{11}+iK_{12})$. 
We can expect that, 
as a result of such an instability, highly-curved vortex filaments 
will be formed. Therefore, the above considered 
"small-curvature" approximation can be valid 
only for finite time. Moreover, it is naive to expect 
a saturation of this instability in a steady-state 
configuration with finite curvature. 
In contrast, we suggest that  frequently reconnection of various parts 
of the filaments, formation of vortex rings etc,  will 
result in persistent
spatio-temporal  dynamics of a highly-entangled vortex state.

In order to follow further development of the instability 
we performed numerical simulations
of the 3D CGLE. We studied a system of $50^3$ dimensionless units 
of Eq. (\ref{cgle1}) 
with no-flux boundary conditions. The numerical solution was performed 
on an R10000 SGI workstation by an
implicit 
Crank-Nicholson algorithm. The number of grid points was 
$100^3$. We performed simulations for $\epsilon=0.02$
in the parameter regime away from amplitude turbulence in 2D
\cite{akw2}. 
As an initial condition we used a straight vortex line perturbed
by  small noise. The results of the simulations are shown in Fig. 1.
We  see that  the length of the vortex line grows. The dynamics seems to
be very rapidly varying in time, and the line intersects itself many times
forming numerous vortex rings. The long-time dynamics show,
however, some kind of saturation when a highly-entangled vortex
state is formed and the length of the line cannot grow anymore due to 
repulsive interaction between closely packed line segments. 
The dependence of the line length on time is 
shown in Fig. 2.
As a measure of the filament length $L$  we used the following quantity:
 $ L \approx S_0^{-1} \int \Theta(A_0 -
|A(x,y,z)|) dx dy dz$, where 
$\Theta $ is a step function: $
\Theta(x)=1$ if $x>0$ and $\Theta=0$ otherwise. 
$A_0 = 0.1$ was used as a threshold value to identify the 
vortex. 
$S_0$ is a constant determined from the condition that for the straight 
line the above integral coincides with the actual length. 
We can identify two distinct stages of the dynamics:
first, fast growth of the length;
second, oscillations of the line's length around some
mean value.  
Remarkably, we observed an increase in the amplitude of 
the 
oscillation moving though  negative $c$ values in the regime of 
spatio-temporal intermittency in the 2D CGLE. 
It is plausible that the symmetry breaking 
mechanism responsible for persistent intermittent 
behavior in the 2D system still has some importance in 3D.

Persistent entangled vortex configurations are known from 
numerical simulations of excitable 
reaction-diffusion systems \cite{karma,winfree}.
Recently, elongation of a vortex filament was also related 
with the meandering instability of a vortex core in 2D
\cite{karma1}. However, there is also a significant difference 
between these two phenomena. For the reaction-diffusion systems 
the meandering instability is typically supercritical, and may
be saturated at a relatively small amplitude of oscillations. 
In application to 3D systems, one expects that 
highly-curved vortex filaments will not develop, and for this 
reason the intersections are very seldom. This picture is 
consistent with the numerical simulations. Consequently, one may
expect that the analysis performed on the basis of a small-curvature 
approximation will be valid for a very long time and can capture 
the actual dynamics of the filament. In the CGLE the 2D
instability 
is subcritical and the velocity of the spiral in an infinite 
system can grow arbitrarily until additional topological defects
nucleate in the wake of the accelerating spiral. In 
the 3D situation we have shown that curvature 
of the filament is even a destabilizing factor. Our numerical simulations
show that the filament does not approach any steady-state. In contrast,
it shows "violent" intermittent behavior, with numerous reconnections and 
splitting of vortex rings. As a result, the approximate equation  
of motion (\ref{accel}) must be considered only as a test for
instability rather than long-time evolution.

In conclusion, we have derived an equation of motion for the vortex filament 
in the CGLE. We have found that in a wide range of parameters of the CGLE
the vortex filament is unstable with respect to spontaneous stretching,
resulting in the formation of persistent entangled vortex configurations. 
This emphasizes the deficiency of previous approaches relating 
local filament velocity to local curvature. 
Our result could be verified in experiments with autocatalytic 
chemical reactions in gels in the regime of oscillatory instability.
Also, the limit of a large cross-diffusion ratio can probably be achieved by 
doping with additional chemicals, thus  changing the relative 
mobility of one chemical 
species with respect to another. 
Recently, the amplitude equation governing the dynamics of an elastic
rod was derived  \cite{tabor}. 
We note that our instability can be formally 
interpreted as 
a dynamics of a thin  rod with {\it negative} elasticity. 
We also speculate that our results are relevant
for inviscid hydrodyanmics. In the limit
of $b,c \to \infty$, Eq. (\ref{cgle}) reduces to the defocusing 
nonlinear Schr\"odinger
equation (NSE), 
which is a paradigm model for compressible inviscid  hydrodynamics. 
Although the vortex lines are stable in the framework of the NSE, the 
corrections arising from the CGLE cause the destabilization
and self-generation of vorticity.

We are grateful to A. Newell, D. Levermore, C. Doering and R. Goldstein 
for illuminating discussions. 
This work was supported by the U.S. Department
of Energy under contracts W-31-109-ENG-38 (IA) and 
ERW-E420  (AB).
The work of IA was also supported by the NSF, 
Office of Science and Technology
Center under contract No. DMR91-20000.

\references{
\bibitem{newell} A.C. Newell, {\it  Envelope Equations}, 
 American Mathematical Society, Providence, RI, 1974 
\bibitem{kuramoto} Y. Kuramoto, {\it Chemical 
Oscillations, Waves and Turbulence}, Springer-Verlag, Berlin, 1983
\bibitem{cross} M. Cross and P.C. Hohenberg, \rmp {\bf 65}
851 (1993)
\bibitem{akw} I.  Aranson, L. Kramer and A. Weber, \pre
{\bf 47}, 3231 (1993); {\it ibid} {\bf 48}, R9 (1993)
\bibitem{akw2}  I.  Aranson, L. Kramer and A. Weber, \prl
{\bf 72}, 2316 (1994)
\bibitem{chate} G. Huber, P. Alstr{\o}m and T. Bohr, \prl
{\bf 69} 2380 (1992) 
H. Chat\^e  and P. Manneville, Physica A
{\bf 224} 348 (1996).
\bibitem{arrechi} F. T.  Arrecchi et al, \prl {\bf 67}, 3749 (1991)
\bibitem{flessel} Q. Ouyang and J.-M.  Flesselles, Nature {\bf 379}, 
6561 (1996) 
\bibitem{perez} M. Ruiz-Villarreal, M.  Gomez-Gesteira and V.  Perez-Villar,
\prl {\bf 78}, 779 (1997)
\bibitem{alt} I. Aranson, H. Levine and L. Tsimring, \prl
{\bf 72}, 2561 (1994)
\bibitem{gold}K.J. Lee, E.C. Cox, and R.E. Goldstein,  \prl
{\bf 76}, 1174 (1996)
\bibitem{frisch} T. Frisch and S. Rica, Physica D  {\bf 61}, 155 (1992)
\bibitem{lmn} J. Lega, J.V. Moloney and A.C. Newell, \prl {\bf 73}, 2978
(1994). 
\bibitem{gog} M. Gabbay, E. Ott, and P. Guzdar, \prl {\bf 78}, 2012 (1997)
\bibitem{karma} F. Fenton and A. Karma, 
 Scroll Wave Dynamics in a Model of Cardiac Tissue with Restitution 
and Fiber Rotation, to be published, 1997
\bibitem{winfree} A. T. Winfree, Physica D {\bf 84}, 126 (1995) 
\bibitem{winfree1} A. Winfree et. al,  CHAOS {\bf 6}, 617 (1996)
\bibitem{keener} J. P.  Keener, Physica D {\bf 31}, 269 (1988)
\bibitem{Hagan} P. Hagan, SIAM J. App. Math. {\bf 42}, 762 (1982) 
\bibitem{karma1} V. Hakim and A. Karma, preprint
\bibitem{tabor} A. Goriely and M. Tabor, \prl {\bf 77}, 3537 (1997)
}

\vspace{.5cm}

\leftline{\hspace{.0cm} \psfig{figure=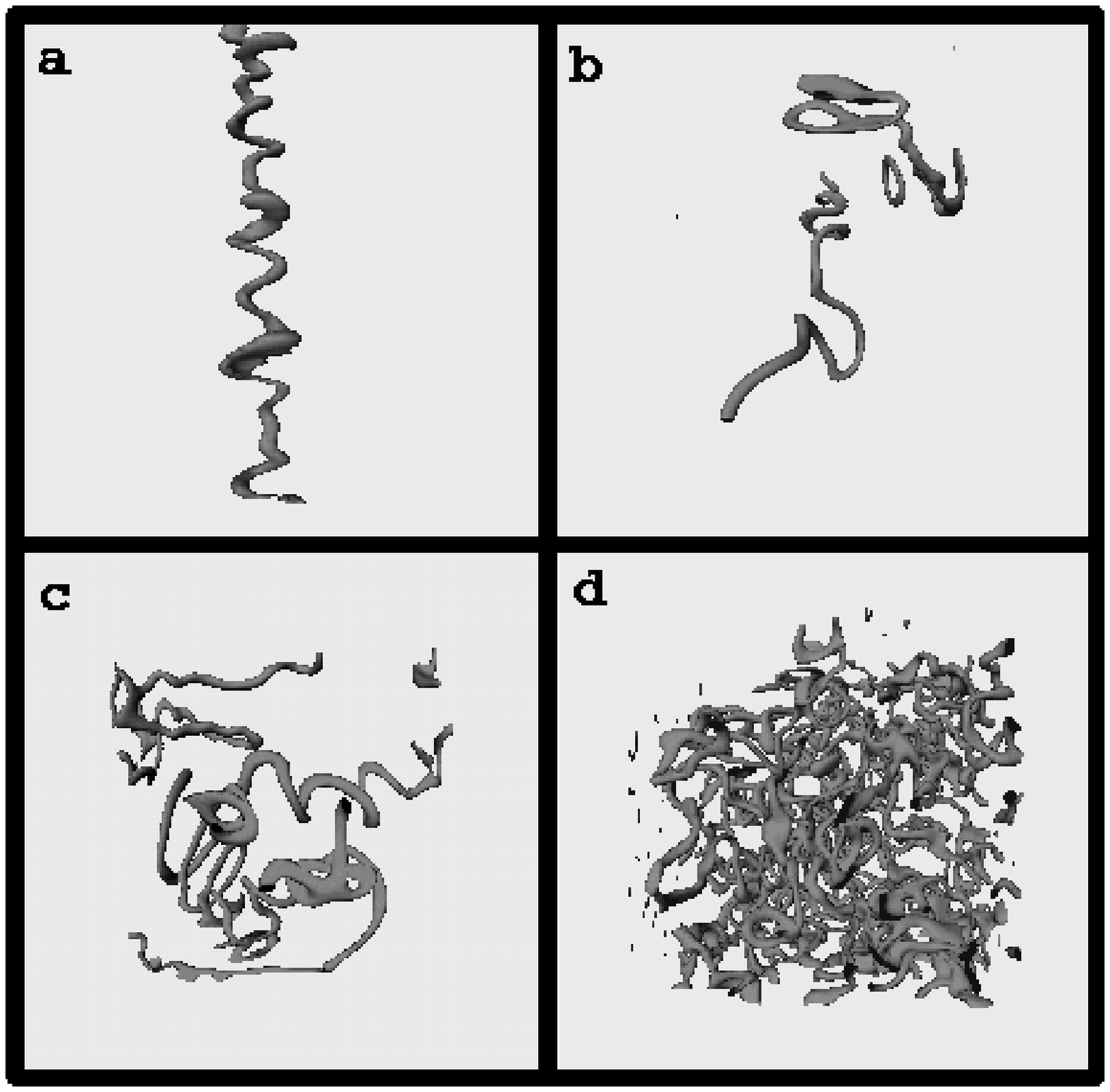,height=3.5in}}
\begin{figure} 
\caption{
Instability of a straight vortex filament. 
3D isosurfaces of $|A(x,y,z)|=0.1$ for 
$\epsilon=0.02$, $c=-0.03 $, shown at  four  
times: 50 (a), 150 (b), 250 (c), 500(d). 
}
\end{figure}

\vspace{-2.5cm}
\leftline{\hspace{.0cm} \psfig{figure=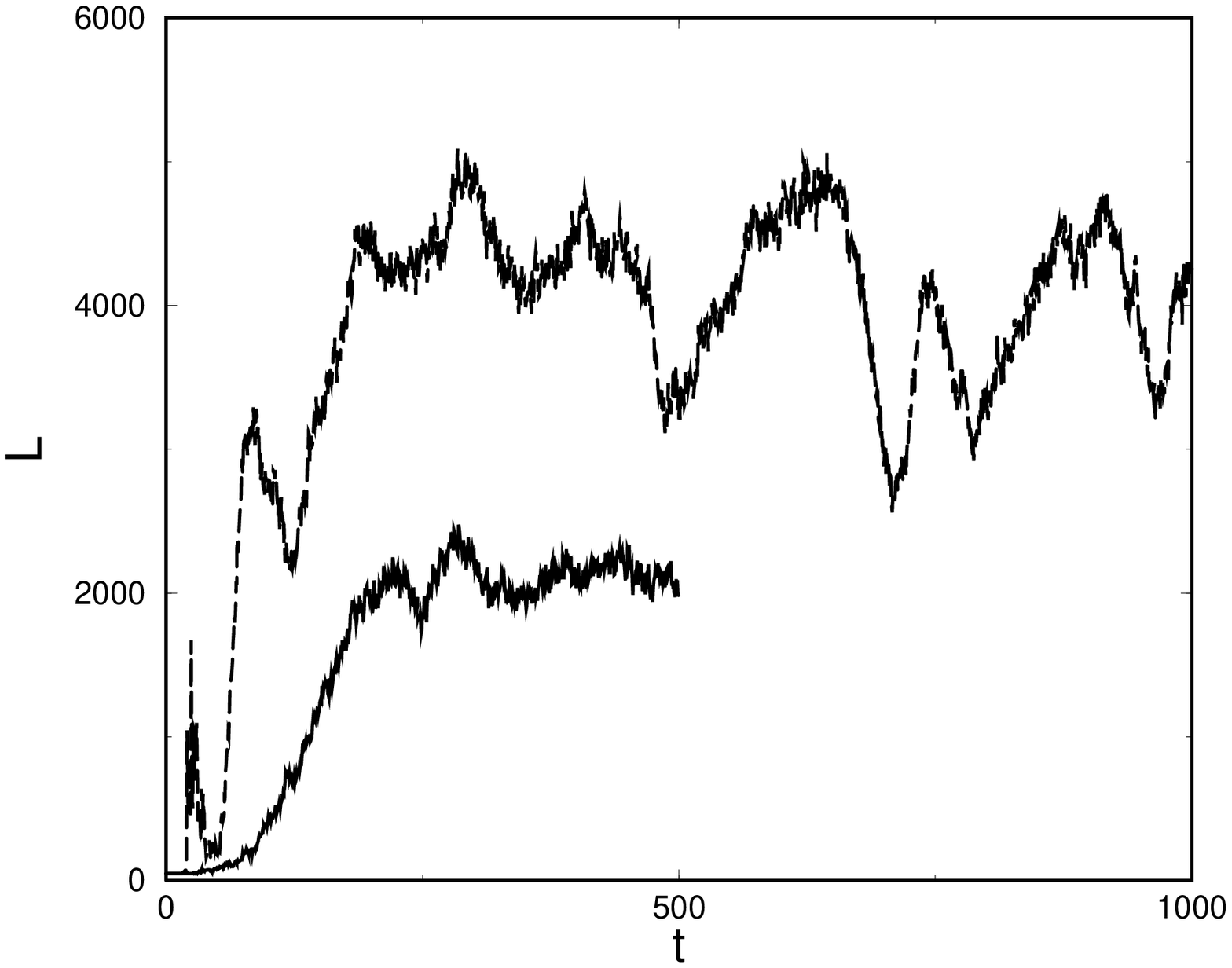,height=3.5in}}
\vspace{-.5cm}
\begin{figure}
\caption{
The dependence of filament length $L$ on time. 
Solid line corresponds to 
$\epsilon=0.02, c=-0.03 $; dashed line corresponds to 
$\epsilon=0.02, c=-0.5$.}
\end{figure}

\end{document}